\documentclass[aps,prl,amsmath,amssymb,reprint,showkeys,showpacs]{revtex4-1}

\newcommand{\nn}{\nonumber}
\newcommand{\ra}{\rightarrow}

\newcommand{\ep}{\epsilon}

\newcommand{\be}{\begin{eqnarray}}
\newcommand{\ee}{\end{eqnarray}}

\newcommand{\bs}{\begin{subequations}\begin{eqnarray}}
\newcommand{\es}{\end{eqnarray}\end{subequations}}
\newcommand{\ba}{\begin{array}}
\newcommand{\ea}{\end{array}}
\newcommand{\bi}{\begin{itemize}}
\newcommand{\ei}{\end{itemize}}
\newcommand{\bei}{\begin{enumerate}}
\newcommand{\eei}{\end{enumerate}}

\usepackage{graphicx}
\usepackage{dcolumn}
\usepackage{bm}

\begin{document} 
 

\title{Space-time dependence of the anomalous exponent of electric transport in the disorder model}

\author{Takeshi Egami}\email{\vspace{-1.5em}egami.takeshi@canon.co.jp}
\author{Koshiro Suzuki} 
\author{Katsuhiro Watanabe}
\affiliation{Analysis Technology Development Center, Canon Inc., 30-2
Shimomaruko 3-chome, Ohta-ku, Tokyo 146-8501, Japan}

\date{\today}
\begin{abstract}
Space-time dependence of the anomalous exponent of electric transport in the disorder model is presented.
We show that the anomalous exponent depends on time, according to the
time-evolution of the 
number of the effective neighbouring sites. 
Transition from subdiffusive to normal transport at long-enough time is recovered.
The above result indicates that the {\it spatial structure}, specifically the {\it
network structure}, of the hopping sites might be a
novel element which determines the anomalous exponent.
This leads to the feature that the scaling property of the electric
transport to time is insensitive to other elements, such as the distance of
the sites or the spatial dimension of the system.
These findings are verified by means of Monte Carlo simulation.
The relation of the result to the conventional knowledge of the Multiple Trapping Model
is shown by deriving it as a special case of the disorder model.
\end{abstract}
\maketitle 
 

\clearpage
\section{Introduction}
Noncrystalline materials, representative examples of which are
organic electro-luminescence, organic photoconductors, and amorphous Si photoconductors,
are readily seen around us.
For instance, photoconductors are used in photocopiers, laser beam printers, and solar cells. 
Since to control the electric conductivity in the above-mentioned devices is crucial for their development and design,
it is necessary to investigate the mechanisms of their carrier transport.
 
Electric conductivity in noncrystalline materials is known to be dominated by the so-called hopping conductance.
It is widely recognized that hopping conductance results in anomalous diffusion (subdiffusion) at a certain mesoscopic scale \cite{Xerox1}.
Experimentally, the anomaly manifests itself in the long-tail of the time-of-flight (TOF) signal \cite{Xerox1, MP, Xerox2}.

Anomalous diffusion has been thoroughly studied in the context of continuous-time random walk (CTRW) \cite{MK}.
In CTRW, the waiting time of the walker $w(t)$ depends on time as $w(t) \sim t^{-(1+\alpha)}$ in the limit $t \ra \infty$, 
where $\alpha$ is called the "anomalous exponent". 
Subdiffusion corresponds to the case $0 < \alpha < 1$, while $\alpha > 1$ describes superdiffusion, or the L\'{e}vy flight.
It is known that the continuum limit of CTRW is described by the fractional diffusion advection equation of order $\alpha$ \cite{MK}.
The crucial feature of CTRW is that the above-mentioned anomaly can be expressed by a single parameter $\alpha$. 
In fact, by varying $\alpha$, the long-tail of the TOF signals can be fitted, as shown in ref. \cite{Xerox1}. 
However, since $\alpha$ is treated as a mere parameter in the CTRW, 
it is significant to study the relation of $\alpha$ and physical quantities of interest, such as the density of state (DOS) of the trap levels, 
or the spatial structure of the hopping sites, for physical understanding of the hopping conductance.

Almost 30 years ago, such a relation was partially given by the "multiple trapping model (MTM)".
In this model, $\alpha$ is related to the DOS of the trap levels and the temperature $T$ as
$
\alpha = T/T_c,
$
where $T_c$ is the typical width of the DOS.
The DOS is assumed to be of the exponential type,
$p(E) = e^{E/(k_B T_c)}/(k_B T_c)$ ($E\leq 0$), where the origin of the energy is
set to the edge of the conduction band.
This type of DOS is typical for disordered irrorganic semiconductors \cite{aSi}.

On the other hand, hopping conductance has been studied mainly by 
two models, the "disorder model \cite{Bassler}", which is studied in this paper, and  
 the "polaron model \cite{SMT}".
The former is suitable for a system where the anomalous charge transport is dominantly
			activated by the static energy disorder of the hopping sites,
			and where the effect of the weak electron-phonon coupling is relatively negligible.
			In contrast, the latter is suitable for a system with 
			strong electron-phonon couplings, and with relatively negligible effects of the
			energy disorder \cite{AEKBNWB}.

In the disorder model, the hopping rate between two sites is described by the classical model of \cite{AHL}, which is based on \cite{MA}. 
As we will review briefly, this model incorporates the information of the spatial structure of the hopping sites, which the MTM does not take into account.
To solve the model of \cite{AHL}, Monte Carlo (MC) simulation has been performed 
\cite{Bassler}, and the relation between the disorder model and the MTM has been studied numerically \cite{HBJK}.
However, the relation of the anomalous exponent and the physical quantities, including the spatial structure of the hopping sites, has not been established theoretically until present. 
Noting that the relation of the anomalous exponent and the width of the DOS has already been established in the MTM,
the theoretical understanding of the "disorder model" seems to be left in a premature state.
The aim of this paper is to promote this theoretical understanding.

In this paper, we first review the disorder model and its related issues.
Secondly, we show the premise and the main result with emphasis on the novel aspects of our study.
Then, the theoretical result is verified by MC simulation.
The methodology of the simulation is briefly explained and the consistency of theory and simulation is exhibited.
Next, the relation of our result and that of the MTM is revealed. 
Finally, we summarize our study.


\section{Theory}
In this study, we consider the disorder model. We adopt as the DOS the exponential type mentioned above, in order for the correspondence to the MTM.
For other types of DOS, e.g. Gaussian DOS, which is typical for organic semiconductors, similar techniques would be applicable.

Before we proceed to the derivation of the anomalous exponent, 
we briefly review the waiting time $w(t)$ of the CTRW and the 
hopping probability of the model of \cite{AHL}. 
In CTRW, the probability distribution that the carrier is at position $x$ at time $t$ is 
given as
\begin{eqnarray}
\rho(x,t)
=
\sum_{x' \neq x} \int_0^t d \tau \rho(x',\tau) \psi(x,x',t-\tau) + \Phi(x,t) \delta_{x0},
\nn
\end{eqnarray}
if the carrier is situated at $x=0$ at $t=0$.
Here, $\Phi(x,t) \equiv 1 - \int_0^t d\tau w(x,\tau)$, and 
$\psi(x',x,t)$ denotes the probability density that the carrier hops from $x$ to $x'$ after waiting time $t$.
The waiting-time probability distribution $w(x,t)$ is given in terms of $\psi(x',x,t)$ as
\begin{eqnarray}
w(x,t)
=
\sum_{x' \neq x} \psi(x',x,t).
\nn
\end{eqnarray}
For the disorder model, $\psi(x',x,t)$ is expressed in the splitted form, $\psi(x',x,t) = w(t) \phi(x',x)$,
where $\phi(x',x)$ is the spatial part of the probability distribution, which is normalized as $\sum_{x'} \phi(x',x) = 1$.
The time-dependent part $w(t)$ is of our interest.

The hopping probability, i.e., the probability per unit time of the carrier to hop from site $i$ to site $j$, which is denoted $\nu_{ij}$, 
is given as \cite{AHL}
\begin{eqnarray}
\nu_{ij}(R_{ij},E_j-E_i)
=
\nu_0 e^{-2\xi R_{ij}-(E_j-E_i)\Theta(E_j-E_i)/k_BT}.
\label{MA}
\end{eqnarray}
Here, $E_i$ is the energy of site $i$, $R_{ij}$ is the distance of site $i$ and $j$,
$\xi$ is the damping factor of the wave function in the localized state, 
$\nu_0$ is a coefficient which represents the magnitude of the hopping rate (typically of the order of $10^{12}$[sec$^{-1}$]), 
$\Theta(x)$ is the Heaviside's step function, and $T$ is the temperature. 
Dependence on microscopic physics, such as the carrier-phonon interaction and the phonon DOS, are assumed to be incorporated in $\nu_0$.
The waiting time of the carrier on site $i$, which we denote $w_i(t)$, is expressed in terms of $\nu_{ij}$ as
\begin{eqnarray}
w_i(t) = \Lambda_i e^{-\Lambda_i t},
\label{wt}
\end{eqnarray}
where $\Lambda_i \equiv \sum_{j\in\mathcal{N}} \nu_{ij}$. 
Here, $\mathcal{N}$ represents the set of {\it all the sites} in the system.
Eq. (\ref{wt}) indicates that the hopping process is modeled as a homogeneous Poisson process with decay rate $\Lambda_i$.
The physical origin of the disorder of energy is that of the spatial
distribution of the hopping sites. 
In the disorder model, the spatial disorder is assumed to be moderate enough
that the positions of the sites fluctuate around the structured lattice
points, according to some distribution (e.g., Gaussian distribution).  
However, for the sake of
simplicity, we incorporate the effect of the spatial disorder to that of the
disorder of the site energy, and assume that the sites compose a structured lattice with
lattice spacing $a$.

Now we calculate the anomalous exponent.
We consider a system with arbitrary spatial dimensions, where the
carrier at each site is allowed to hop to all the sites in the system. 
The crucial observation is that the waiting-time distribution Eq. (\ref{wt}) is merely valid for a certain energy.
If the energy is distributed according to some probability distribution, then the ensemble average with respect to the energy distribution,
which we denote $\langle w(t) \rangle$, should be regarded as the effective waiting time of the system \cite{Kivelson}. 
\begin{eqnarray}
&&\langle w(t) \rangle=
	\left(\prod_{j\in \mathcal{N}}\int_{-\infty}^\infty d \epsilon_{ij}
	p_L(\epsilon_{ij})
	\right)
	w_i(t)
\nn
\\&&=
	\left(\prod_{j\in \mathcal{N}}\int_{-\infty}^\infty\!\! d \epsilon_{ij}
	\frac{e^{-|\ep_{ji}|/k_BT_c}}{2k_BT_c}
	\right)\!\!
\nn\\
&&\times
	\left(
			\sum_{k\in\mathcal{N}}\!\! K_{ki}
			e^{
				-\frac{\epsilon_{ki}\Theta(\epsilon_{ki})}{k_BT}}
			e^{
				-\sum_{l\in\mathcal{N}} K_{li}
			e^{-\epsilon_{li}\Theta(\epsilon_{li})/k_BT} t
			}
	\right),
	\label{wtev}
\end{eqnarray}
where
\begin{eqnarray}
\vspace{-1.0em}
K_{ji}\equiv \nu_0e^{-2\xi R_{ji}} 
\label{defK}
\end{eqnarray}
, and $\ep_{ji}\equiv E_{j} - E_i$ is the difference of the energy of site
$i$ and $j$. 
Note that translational invariance is assumed, as a result of the
spatial coarse-graining due to the integration with respect to site
energies, and the subscript $i$ is omitted in the expression $\langle w(t) \rangle$.
In Eq. (\ref{wtev}), we have utilized the fact that the energy difference $\ep_{ji}$, which is the difference of two identical probabilistic variables, obeys the Laplace distribution $p_L(\ep_{ji}) \equiv e^{-|\ep_{ji}|/k_BT_c}/(2k_BT_c)$.
We can calculate the integrals in Eq. (\ref{wtev}) by two successive transformations of the integration variables, i.e., 
$A_{ji}=e^{-\ep_{ji} /k_B T}$ for the first step, 
and $C_{ji}=K_{ji} A_{ji} t$ for the second \cite{US}.
By the first transformation $A_{ji}=e^{-\ep_{ji} /k_B T}$, and using $d A_{ji}/d \epsilon_{ji} =-A_{ji}/{k_BT}$, we obtain
\be
&&
		\int_{0}^\infty d \epsilon_{ji}
		e^{-K_{ji}e^{-\epsilon_{ji}/k_BT}t}
		\frac{e^{-|\epsilon_{ji}|/k_BT_c}}{2k_BT_c}
	\nn\\&&=
		\frac{1}{2}
		\frac{T}{T_c}\int_{0}^1 dA_{ji} 
		A_{ji}^{-1+\frac{T}{T_c}}
		e^{-K_{ji}A_{ji}t}.
\ee
Then, by the second transformation $C_{ji}=K_{ji} A_{ji} t$, we obtain
\be
	&&
	\frac{1}{2}
	\frac{T}{T_c}\int_{0}^1 dA_{ji} 
		A_{ji}^{-1+\frac{T}{T_c}}
		e^{-K_{ji}A_{ji}t}
	\nn\\&&=
		\frac{1}{2}
		\frac{T}{T_c}(K_{ji}t)^{-\frac{T}{T_c}}\int_{0}^{K_{ji}t} dC_{ji}
		C^{-1+\frac{T}{T_c}}_{ji}
		e^{-C_{ji}}
	\nn\\&&=
	\frac{1}{2}
		\frac{T}{T_c}(K_{ji}t)^{-\frac{T}{T_c}}
		\gamma\left(\frac{T}{T_c},K_{ij}t\right).
		\label{11072901}
\ee		
Here, 
\be
	\gamma(T/T_c,K_{ij}t)\equiv \int_0^{K_{ij}t}d \tau
		\tau^{-1+T/T_c}e^{-\tau}
	\label{incompgamma}
\ee
 is the lower incomplete gamma function.
Similarly, we obtain
\be	
	&&
		\int_{0}^\infty d \epsilon_{ji}
		e^{-\epsilon_{ji}/k_BT}
		e^{-K_{ji}e^{-\epsilon_{ji}/k_BT}t}
		\frac{e^{-|\epsilon_{ji}|/k_BT_c}}{(2k_BT_c)}
	\nn\\&&=
	\frac{1}{2}
		\frac{T}{T_c}(K_{ji}t)^{-1-\frac{T}{T_c}}
		\gamma\left(\frac{T}{T_c}+1,K_{ij}t\right).
		\label{11072902}
\ee		
Substituting Eqs. (\ref{11072901}) and (\ref{11072902}) into
Eq. (\ref{wtev}), we have
\begin{eqnarray}
\begin{array}{rl}
\langle w(t) \rangle
&=
\sum_{k\in \mathcal{N}}
\left\{
\frac{K_{ik}}{2}
\left[	
	\frac{T}{T_c}\frac{\gamma\left(\frac{T}{T_c}+1,K_{ik}t\right)}{(K_{ik}t)^{1+T/T_c}}
	+e^{-K_{ik}t}
\right]
\right.
	 \\
&
\times
\prod_{j\in \mathcal{N},j\neq k}
\left.
\frac{1}{2}\left[
	\frac{T}{T_c}\frac{\gamma\left(\frac{T}{T_c},K_{ij}t\right)}{(K_{ij}t)^{T/T_c}}
	+e^{-K_{ij}t}
\right]
\right\}.
\end{array}
\label{11060902}
\end{eqnarray}

It is important that the following relations hold for the sites which satisfy
 the condition $K_{ik}t\ll1$,
 \be
&&
	\frac{1}{2}
	\left[
		\frac{T}{T_c}\frac{\gamma\left(\frac{T}{T_c},K_{ik}t\right)}{(K_{ik}t)^{T/T_c}}
		+e^{-K_{ik}t}
	\right]\simeq 1,
	\label{11071502}
\\
&&
	\frac{1}{2}
	\left[	
		\frac{T}{T_c}\frac{\gamma\left(\frac{T}{T_c}+1,K_{ik}t\right)}{(K_{ik}t)^{1+T/T_c}}
		+e^{-K_{ik}t}
	\right]\simeq 1.
	\label{11071503}
\ee
Furthermore, for the sites which also satisfy the condition $K_{ik}\ll1$, we obtain
\be
	\frac{K_{ik}}{2}
	\left[	
		\frac{T}{T_c}\frac{\gamma\left(\frac{T}{T_c}+1,K_{ik}t\right)}{(K_{ik}t)^{1+T/T_c}}
		+e^{-K_{ik}t}
	\right]
	\simeq 0,
	\label{11071504}
\ee
in combination with Eq. (\ref{11071503}). 
Therefore, the sites which satisfy $K_{ik}t\ll1$ and $K_{ik}\ll1$ are irrelevant to the effective waiting time. 
Then we can rewrite Eq. (\ref{11060902}) as 
\be
	\langle w(t) \rangle
&\simeq&
	\sum_{k\in \mathcal{N}_{R}(t)}
	\left\{
	\frac{K_{ik}}{2}
	\left[	
		\frac{T}{T_c}\frac{\gamma\left(\frac{T}{T_c}+1,K_{ik}t\right)}{(K_{ik}t)^{1+T/T_c}}
		+e^{-K_{ik}t}
	\right]
	\right.
\nn\\&	\times&
	\prod_{j\in \mathcal{N}_{R}(t),j\neq k}
	\left.
	\frac{1}{2}\left[
		\frac{T}{T_c}\frac{\gamma\left(\frac{T}{T_c},K_{ij}t\right)}{(K_{ij}t)^{T/T_c}}
		+e^{-K_{ij}t}
	\right]
	\right\}.
	\label{11071501}
\ee
Here, $\mathcal{N}_{R}(t)$ is the set of the "effective neighbours" at time $t$, 
which is defined by removing the sites which satisfy $K_{ik}t\ll1$ and $K_{ik}\ll1$ from $\mathcal{N}$.

Next, to obtain further an approximate analytic expression for the effective
waiting time, we attempt to approximate the lower incomplete gamma
function by the gamma function.
From Eq. (\ref{incompgamma}), it is obvious that 
\begin{eqnarray}
\gamma(T/T_c, K_{ij}t) \simeq \Gamma(T/T_c)
\label{approx}
\end{eqnarray}
for $K_{ij}t \gg 1$.
To be more specific, we introduce a criteria which defines the precision
of the approximation Eq. (\ref{approx}).
We consider a constant $C_\gamma$, where the following
replacement is performed for the case $K_{ij}t > C_\gamma$:
\begin{eqnarray}
&&
\gamma\left(T/T_c,K_{ij}t \right)
\ra
\Gamma\left(T/T_c\right)
\hspace{1em}
(K_{ij}t>C_\gamma, 0< T/T_c \le 2).
\nn
\end{eqnarray}
In other words, $C_\gamma$ defines the precision of the approximation 
Eq. (\ref{approx}).
Then, we can define the critical radius $R(t)$ at time $t$,
\be
	R(t)\equiv -\frac{\ln\frac{C_\gamma}{\nu_0t}}{2\xi}.
	\label{11071505}
\ee
For sites with $R_{ij}<R(t)$, approximation Eq. (\ref{approx}) is applied in Eq. (\ref{11071501}).

We approximate Eq. (\ref{11071501}) by only including the neighbouring sites 
with $K_{ij}t > C_\gamma$.
Then,
\begin{eqnarray}
\hspace{-0.5cm}
\begin{array}{rl}
	\langle w(t) \rangle
&\!\!\simeq
\sum_{k\in \mathcal{N}_{R}'(t)}
\left\{
\frac{K_{ik}}{2}
\!\!\left[	
	\left(\frac{T}{T_c}\right)^2\!\!\frac{\Gamma\left(\frac{T}{T_c}\right)}{(K_{ik}t)^{1+T/T_c}}\!
	+\!e^{-K_{ik}t}
\right]
\right.
	 \\
&\!\!
\times
\prod_{j\in \mathcal{N}_{R}'(t),j\neq k}
\left.
\frac{1}{2}\left[
	\frac{T}{T_c}\frac{\Gamma\left(\frac{T}{T_c}\right)}{(K_{ij}t)^{T/T_c}}
	+e^{-K_{ij}t}
\right]
\right\},
\end{array}
\label{11061201}
\end{eqnarray}
where $\mathcal{N}_R'(t)$ is defined as a set of neighbouring sites 
with $R_{ij}<R(t)$.
Let us denote the set of the $n$ th-nearest neighbouring sites as
$\mathcal{N}_n$.
From the definition of $K_{ij}$, Eq. (\ref{defK}), it is obvious that 
$K_{ik}t = e^{2\xi ma} K_{ij}t$ ($k \in \mathcal{N}_{n-m}$, 
$j \in \mathcal{N}_n$, $m \in \bf{N}$) holds for the $n$ th-nearest and
the $(n-m)$ th-nearest neighbours.
For a system with typical parameters (e.g., the case shown in the
numerical simulation later), $e^{2\xi a} \gg 1$ holds.
Hence, for time $t$ where the $n$ th-nearest neighbours are excluded in
Eq. (\ref{11071501}), i.e., $K_{ij}t \ll 1$ ($j \in \mathcal{N}_n$),
there exists $m \in \bf{N}$ such that $K_{ij}t > C_\gamma$ 
($j \in \mathcal{N}_{n-m}$), i.e., $(n-m)$ th-nearest neighbours are
included in $\mathcal{N}_R'(t)$.
For the case $m=1$, $(n-1)$ th-nearest neighbours
are included in $\mathcal{N}_R'(t)$, and hence 
$\mathcal{N}_R(t)$ in Eq. (\ref{11071501}) and $\mathcal{N}_R'(t)$ in
Eq. (\ref{11061201}) coincide.
In general, the $(n-m)$ th-nearest neighbours with $1 < K_{ij}t <
C_\gamma$ is omitted in Eq. (\ref{11061201}), but as far as 
$e^{2\xi a} \gg 1$ holds, their effect on the
effective waiting time is subdominant compared to that of the sites
with $R_{ij} < R(t)$.

We can introduce a characteristic time $\tau_n$, at which the
 $n$ th-nearest neighbours contribute to the effective waiting time, as
\be
	\tau_n\equiv\frac{C_\gamma}{\nu_0e^{-2\xi a n }}.
	\label{11061501}
\ee
From Eq. (\ref{11061501}), we obtain 
$\tau_{n+1}=e^{2\xi a  }\tau_n$. 
As mentioned above, $\tau_{n+1} \gg \tau_n$ holds for typical systems. 
This indicates that the $(n+1)$ th-nearest neighbours are negligible for
time scales around or below $\tau_n$.
Thus, we can define a time interval $I_N$ where the number of the
"effective neighbors" equals $N$.
Then, it can be stated that Eq. (\ref{11061201}) is valid for $t \in
I_{N_R(t)}$,
where $N_R(t)$ is the number of sites in $\mathcal{N}_R'(t)$.
From Eq. (\ref{11071505}), the critical radius $R(t)$ increases
monotonically with time. 
As a result, $N_{R}(t)$ increases monotonically
as $\lim_{t\rightarrow\infty}N_{R}(t) = N_{tot}$, where $N_{tot}$ is
the total number of sites in the system.

From Eq. (\ref{11061201}), we can derive simple results for the following two limiting cases: $N_R(t)T \ll T_c$ and $N_R(t)T \gg T_c$.
For the case $N_R(t)T \gg T_c$, taking the limit $T \ra \infty$ with $T_c$ fixed in Eq. (\ref{11061201}) leads to
\begin{eqnarray}
	\langle w(t) \rangle
\simeq
\frac{1}{2^{N_R}}
\sum_{ k\in \mathcal{N}_{R}'(t)}
K_{ik}e^{-\sum_{ j\in \mathcal{N}_{R}'(t)}K_{ij}t}
.
\label{11061202}
\end{eqnarray}
This is the usual Poisson distribution, and hence $\alpha = 1$, i.e.,
the diffusion is normal. 
This is consistent with our understanding that the band conduction dominates at high temperatures.

For the case $N_R(t)T \ll T_c$, one can see that the first term is
 dominant, respectively, in the two squared brackets in Eq. (\ref{11061201}):
\begin{eqnarray}
\begin{array}{rl}
	\!\!\langle w(t) \rangle
\!\!\!\!&\simeq\!
\sum_{ k\in \mathcal{N}_{R}'(t)}
\left\{\!
\frac{K_{ik}}{2}
	\left(\frac{T}{T_c}\right)^2\Gamma\left(\frac{T}{T_c}\right)\!(K_{ik}t)^{-\left(\frac{T}{T_c}+1\right)}
\right.
	 \\
&\times
\prod_{ j\in \mathcal{N}_{R}'(t),j\neq k}\!\!
\left.
\left(
\frac{1}{2}
	\frac{T}{T_c}\Gamma\left(\frac{T}{T_c}\right)\!(K_{ij}t)^{-\frac{T}{T_c}}
\right)	\!
\right\}
\\
&
\propto t^{-1-N_R(t)\frac{T}{T_c}}.
\end{array}
\label{11061203}
\end{eqnarray}
The asymptotic form $\langle w(t) \rangle \propto t^{-(1+N_R(t)T/T_c)}$ ($t
\in I_{N_R(t)}$) indicates that the coarse-grained hopping conductance
possesses the feature of subdiffusion with a time-dependent anomalous exponent $\alpha(t) = N_R(t)T/T_c$.

To summarize, we derived the asymptotic form of the anomalous exponent,
\begin{eqnarray}
\alpha(t)	
&=&
\left\{
\begin{array}{cl}
{N}_{R}(t)\frac{T}{T_c}	& ({N}_{R}(t)T\ll T_c)\\
1				& ({N}_{R}(t)T\gg T_c)
\end{array}
\right.
,\label{anom-exp-N3}
\end{eqnarray}
which is valid for time $t \in I_{N_R(t)}$.
Eq. (\ref{anom-exp-N3}) indicates that the anomalous exponent does
depend on the {\it spatial structure} of the hopping sites.
Moreover, an interesting feature of Eq. (\ref{anom-exp-N3}) is that 
the {\it network structure} of the hopping sites is the only spatial element 
which contributes to the anomalous exponent. 
In other words, other elements, such as the distance between the
sites and the spatial dimension of the system, are insensitive
to the time-dependence of the electric transport, as
long as $N_R(t)$ is identical.
It is also clear from Eq. (\ref{anom-exp-N3}) that the anomalous exponent 
depends on time. 
Since ${N}_{R}(t)$ increases with time,
$\alpha(t)$ eventually becomes $1$ at some time in the future, i.e., 
subdiffusive transport necessarily becomes normal. 
This is consistent with the common knowledge of nonequilibrium
statistical mechanics that the Markovian property is recovered by a
proper coarse graining \cite{Kub2}.
A novel indication of the above result is that, since $N_R(t)$ increases
with time, a discrepancy with the MTM, which corresponds to the case of $N_R(t)=1$, shows up.
This issue will be discussed later.

\section{Simulation}
The theoretical asymptotic form of the anomalous exponent
 Eq. (\ref{anom-exp-N3}) is verified by the MC simulation of the hopping conductance.
We mainly focus to the following three theoretical predictions :
(i) the transition from the subdiffusive transport to the normal transport, 
(ii) the transient behavior of the anomalous exponent, 
and (iii) the effect of the network structure of the hopping sites to the anomalous exponent.

The anomalous exponent results in the time-dependence of the transport coefficients.
This fact suggests that one cannot adopt the naive definition of the transport coefficients, e.g., 
$D = \left[\langle x^2(t_{tot}) \rangle - \langle x(t_{tot}) \rangle^2 \right]^{\frac{1}{2}}/t_{tot}$ for the diffusion coefficient 
\cite{Bassler}, where $t_{tot}$ is the total duration time of the hopping process.
Rather, one should estimate the corresponding moments at {\it intermediate} times, e.g.,
\begin{eqnarray}
D(t) = \left[ \langle x^2(t) \rangle - \langle x(t) \rangle^2 \right]^{\frac{1}{2}}/t 
\hspace{0.5em} (0 \leq ^{\forall} t \leq t_{tot}).
\label{D}
\end{eqnarray}
Estimate such as Eq.  (\ref{D}) cannot be performed in the conventional MC methods \cite{Bassler},
which is difficult to synchronize the hopping procedure of the independent carriers.
An algorithm is developed to synchronize the hopping of the carriers, whose details will be published elsewhere \cite{Watanabe}.
We adopted this algorithm to calculate the time-dependence of the transport coefficients, or, equivalently, the anomalous exponent.
To be more precise, in order to estimate $\alpha$ from the MC simulation results, 
we used the following relation which holds for subdiffusion: 
\begin{eqnarray}
 \langle x^2(t) \rangle - \langle x(t) \rangle^2 \propto t^{\alpha}
\hspace{0.5em} (0 \leq ^{\forall} t \leq t_{tot}).
\label{x2}
\end{eqnarray}
From Eq. (\ref{x2}), we can estimate $\alpha$ by fitting the function
$f(t)=Ft^{\alpha}$, where $F$ is a constant, to the simulation result at time $t$. 
This estimate requires the synchronization of the hopping carriers.

Before proceeding to discuss the simulation results,  
we estimate the value of $C_\gamma$ in Eq. (\ref{11071505}). 
Since the incomplete gamma function $\gamma\left(T/T_c,C_\gamma\right)$
asymptotically approaches to the gamma function $\Gamma\left(T/T_c\right)$
, at least in $6$--$7$ digits accuracy, 
 at $C_\gamma=100$, we chose $C_\gamma=100$ for the case $0.05\le T/T_c\le 1$.
Substituting typical values $t=10^{-2}$ [sec],
$2\xi a=10$, $a=10^{-9}$ [m] and $\nu_0=10^{12}$ [$\mathrm{sec^{-1}}$]
 to Eq. (\ref{11071505}),
 we obtain the critical radius $R=2.8\times10^{-9}$ [m] for $C_\gamma =
 100$. 
This indicates that it is enough 
to consider at most the third-nearest neighbours for typical systems. 
Therefore, in realistic time scales, the number of the effective neighbours is
finite, typically at most of the order of ten. 
This is consistent with common hopping models, which assume that the
number of the neighbouring sites to which hopping is allowed is finite \cite{Bassler}.
In addition, using the above parameter values, we obtain a relation of
characteristic times, $\tau_{n+1}=e^{2\xi a  }\tau_n\simeq 2.2\times 10^4\tau_n$. 
As mentioned in the previous section, this confirms that
$(n+1)$th-nearest neighbours are negligible for time scales around or below $\tau_n$.

Now, we verify the validation of Eq. (\ref{anom-exp-N3}) by the MC simulation.
First, in order to illuminate the transition from subdiffusive to normal
transport,
we chose a one-dimensional system with carriers allowed to hop up to the second-nearest neighbouring sites. 
The condition of the simulation is as follows;
the number of carriers is $N_P=1000000$,
the parameters are $2\xi a = 10 $, $\nu_0 = 10^{12}$[sec$^{-1}$], $T /T_c = 0.375$. 
These values correspond to the characteristic times $\tau_1=2.2\times 10^{-6}$[sec],  $\tau_2=4.9\times 10^{-2}$[sec], etc.
Initially, all the carriers are rested at the origin.
For the case where the correlation between the carriers can be neglected, this initial configuration is sensible,
since in the novel algorithm \cite{Watanabe} neighbours are generated independently for each carrier,
where each neighbour spans a probabilistically independent sample space.

We can theoretically estimate characteristic values of $\alpha(t)$ at characteristic times.
Since $T /T_c = 0.375$ and there are two sites in the 1st and
2nd-nearest neighbours in the system, respectively, $\alpha(t)$ becomes as follows:
\be
	\alpha(t)
	\simeq
	\left\{
	\ba{cc}
		0.75	&	(t=\tau_1) \\
		1.0	&	(t\geq \tau_2)
	\ea
	\right..
	\label{11080101}
\ee
Eqs. (\ref{11080101}) indicate that $\alpha(t)$ approaches 1.0 eventually,
taking the value 0.75 at $t=\tau_1$.

\begin{figure}[htb]
\includegraphics[width=8.5cm]{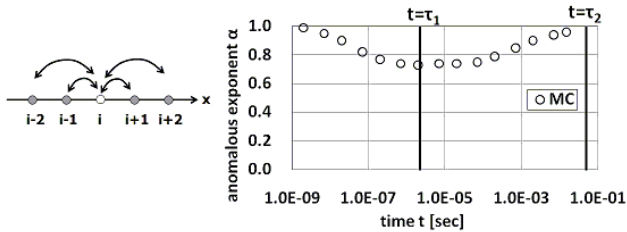}	
\caption{The result for the MC simulation of the transition from
 subdiffusive to normal transport. 
The system is one-dimensional with hopping up to the second-nearest neighbours is allowed. 
}
\label{fig-normal}
\end{figure}

The result of the MC simulation is shown in Fig. \ref{fig-normal}.
One can see that the result is compatible with the theoretical estimate,
Eqs. (\ref{11080101}). 
Thus, transition from subdiffusive to normal transport is verified. 
Moreover, it can be seen that the transport is normal 
in the early stage of the diffusion. 
In this stage, since only carriers with short life-times can hop to the
neighbouring sites, the hopping is approximately unimodal.
Therefore, the transport is almost normal.

%

Next, in order to illuminate the transient behavior of the anomalous exponent, 
we chose a one-dimensional system with carriers allowed to hop up to the third-nearest neighbouring sites. 
The condition of the simulation is as follows;
the number of carriers is $N_P=1000000$,
the parameters are $2\xi a = 10 $, $\nu_0 = 10^{12}$[sec$^{-1}$], 
$T /T_c = 0.1$.
These values correspond to the characteristic times $\tau_1=2.2\times
10^{-6}$[sec],  
$\tau_2=4.9\times 10^{-2}$[sec], $\tau_3=1.1\times 10^{3}$[sec], etc.

Characteristic values of $\alpha(t)$ at characteristic times can be
estimated theoretically. 
Since $T /T_c = 0.1$ and there are two sites in the 1st, 2nd and
3rd-nearest neighbours in the system, respectively,
$\alpha(t)$ becomes as follows:
\be
	\alpha(t)
	\simeq
	\left\{
	\ba{cc}
		0.2	&	(t=\tau_1)\\
		0.4	&	(t=\tau_2)\\
		0.6	&	(t=\tau_3)
	\ea
	\right..
	\label{11061601}
\ee

\begin{figure}[htb]
\includegraphics[width=8.5cm]{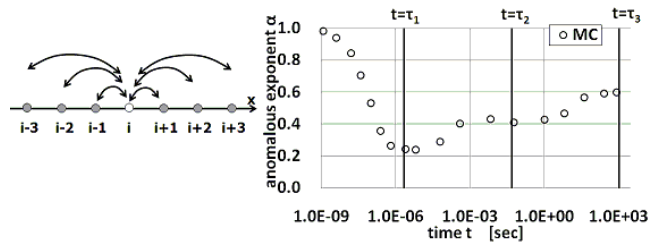}	
\caption{The result for the MC simulation of
the transient behavior of the anomalous exponent.
The system is one-dimensional with hopping up to the third-nearest neighbours is allowed. }
\label{fig1}
\end{figure}

The result of the MC simulation is shown in Fig. \ref{fig1}.
One can see that the result is compatible with the theoretical estimate, Eqs. (\ref{11061601}). 
In addition, plateaus are found around each characteristic time $\tau_n$
($n=1,2,3$). 
This indicates that $\alpha(t)$ can be regarded as constant at a certain
time interval.  
This is compatible with the common treatment that $\alpha(t)$ is regarded as constant \cite{Xerox1}.
%

\begin{figure}[htb]
\includegraphics[width=8.5cm]{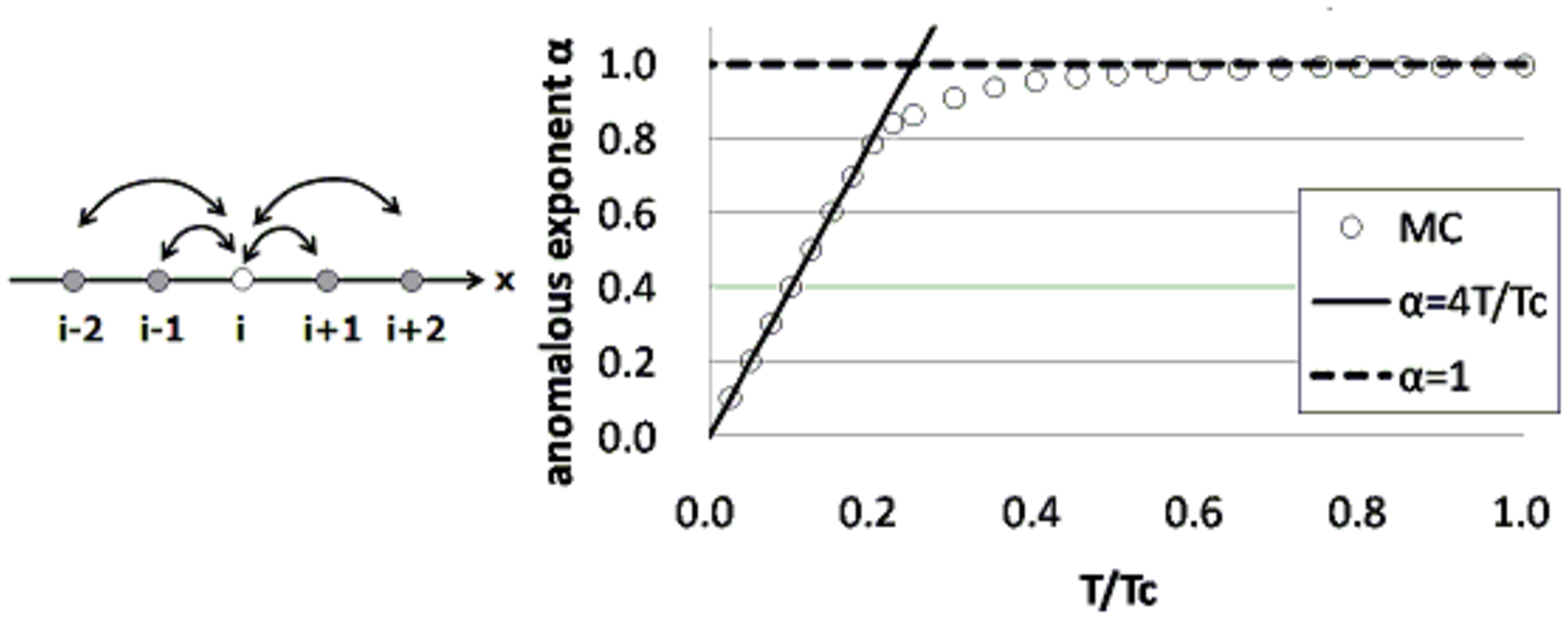}	
\includegraphics[width=8.5cm]{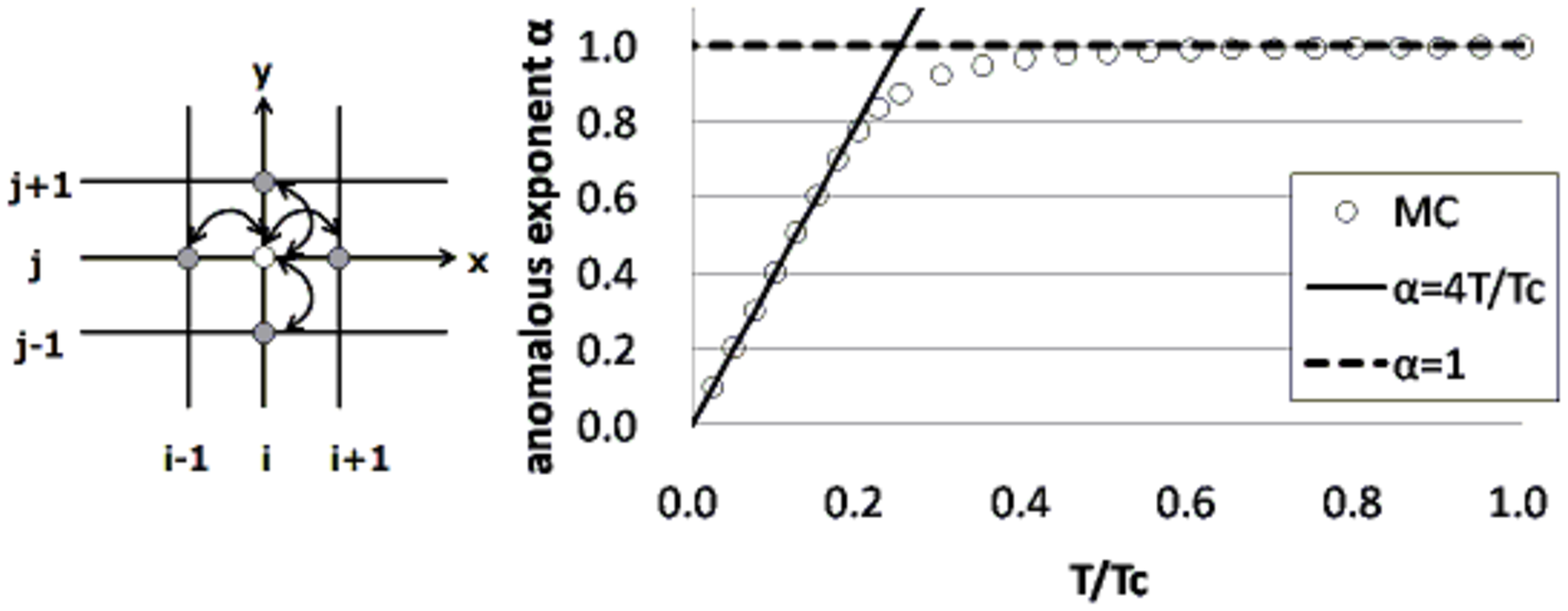}
\caption{Comparison of the theoretical and the simulational results
for the relation between the anomalous exponent $\alpha$ and $T/T_c$.
The upper figure is for a one-dimensional system with hopping up to the second-nearest neighbours are allowed 
($N_R(\infty)=4$), 
while the lower figure is for a two-dimensional system with that only to the nearest neighbours are allowed 
($N_R(\infty)=4$).
The network structure of the two systems are identical (homeomorphic), while other features such as the distance of the 
neighbouring sites and the spatial dimension of the system are different.}
\label{fig}
\end{figure}

Finally, in order to illuminate the effect of the {\it network structure} of the
hopping sites, 
two cases are chosen for illustration.
One is a one-dimensional system with carriers allowed to hop up to the second-nearest neighbours
(4 sites in total, see the upper figure in Fig. \ref{fig}).
Another is a two-dimensional system with carriers allowed to hop only to the nearest neighbours
(4 sites in total, see the lower figure in Fig. \ref{fig}).
The condition of the simulation is as follows;
the number of carriers is $N_P=1000000$,
the parameters are $\xi a = 1$, $\nu_0 = 10^{12}$[sec$^{-1}$], and 
$T/T_c$ is varied in the range 0.025 - 1.0. 
The duration of the simulation is chosen to be larger than $\tau_2$.
From Eq. (\ref{anom-exp-N3}), 
it is expected theoretically that the results for the two cases are identical.

The result of the MC simulation is shown in Fig. \ref{fig}.
One can see that the results for the two cases are identical.
For both cases, the simulation result coincides with the theoretically derived line $\alpha = 4 T/T_c$ for $T/T_c < 0.2$,
and it reaches $\alpha = 1$ for $T/T_c > 0.6$.
The {\it network structures} of the two systems are identical, i.e., they are homeomorphic, 
while other features such as the distance between the sites and the spatial dimension differ.
This result indicates that the anomalous exponent is determined by the {\it network structure} of the hopping sites.

The origin of this network structure dependence resides in the feature of the model that
the probability distribution of the waiting time is described by a
Poisson distribution Eq. (\ref{wt}) for a given site energy.
Due to this feature, the dependence of 
the time-dependent part of
$\langle w(t) \rangle$ on the distance of the sites appear 
(through $K_{ij}t$)
in the upper limit of the integral of the incomplete gamma function, 
which is wiped out to infinity in the long-time limit,
leaving the power-law dependence of $K_{ij}t$.
The emergence of the Poisson distribution is attributed to the independence of each hopping procedure, which is valid 
in the weak-coupling limit of the carriers.
For the strong-coupling case, intricate distribution will possibly emerge, and it is beyond the scope of the present paper.

\section{Discussion}
One of the major results of this paper is that the anomalous exponent
in the disorder model \cite{AHL} is in general larger than that in the
MTM \cite{MTM}.
To the extent of our knowledge, an exact correspondence between the disorder model and the MTM has not been stated clearly.
In the remainder we will state their correspondence.

In the MTM, the carriers reside in the infinite tower of
almost-continuous energy levels ("trapped levels"),
as well as in the conduction band.
The carriers in the conduction band ("conduction carriers") and those in the trapped level ("trapped carriers") are treated separately.
The "conduction carriers" are allowed to move to the neighbouring
sites according to normal diffusion, while there is no diffusion for the ``trapped carriers''.

On the other hand, there is no distinction between the "conduction carriers" and the "trapped carriers" 
in the disorder model.
In particular, the "trapped carriers" are allowed to hop to the
neighbouring sites as well. 

These two models apparently seem independent, but
by imposing a constraint on the disorder model that hopping to
the neighbouring sites are allowed {\it only when the energy level of
the sites coincide with} $E_c$ (the energy of the lower edge of the
conduction band), the anomalous exponent of the MTM can be obtained. 
To illuminate this correspondence, we chose a one-dimensional system
with a carrier allowed to hop only to the nearest neighbouring sites. 
Imposing the above constraint to Eq. (\ref{MA}), the hopping rate from site $i$ to site $i\pm1$ becomes
\begin{eqnarray}
	\nu_{i,i\pm1}'
	=
	\left\{
		\begin{array}{cc}
		\nu_{1}e^{-(E_c-E_i)/k_BT}&(E_{i\pm1}=E_c)\\
		0		          &(E_{i\pm1}<E_c)
		\end{array}
	\right.		.
	\label{11080401}
\end{eqnarray}
Here, $\nu_1\equiv\nu_0e^{-2\xi a}$.
Using Eq. (\ref{11080401}) and choosing $E_c=0$ for simplicity, we
obtain the waiting time for this case $w'$ as follows:
\be
	\!\!\!\!&&\!w'(E_{i-1},E_i,E_{i+1})\nn
\\	
\!\!\!\!&&\!=
	\left\{\!
		\begin{array}{ll}
			2\nu_{1}e^{E_i/k_BT}
			e^{-2\nu_{1}e^{E_i/k_BT}t}&\!\!
				{\small(E_{i+1}=0,\  E_{i-1}=0)}
		\\
		\nu_{1}e^{E_i/kT}
		e^{-\nu_{1}e^{E_i/k_BT}
		t}
		&\!{\small(E_{i+1}=0,\  E_{i-1}<0)}\\
		\nu_{1}e^{E_i/k_BT}
		e^{-\nu_{1}e^{E_i/k_BT}
		t}
		&\!{\small(	E_{i+1}<0,\  E_{i-1}=0)}\\
		0	&\!\!{\small(E_{i+1}<0,\  E_{i-1}<0)}\\
		\end{array}
	\right.				\!\!\!.
	\label{11080201}
\ee
By averaging over the energy, we can obtain the effective waiting time as
\be
&&
		\langle w'(E_{i-1},E_i,E_{i+1})\rangle
	\nn\\	
	&=&	\int_{-\infty}^{0}d \left(\frac{E_{i+1}}{k_BT_c}\right)
		\int_{-\infty}^{0}d \left(\frac{E_{i}}{k_BT_c}\right)
		\int_{-\infty}^{0}d \left(\frac{E_{i-1}}{k_BT_c}\right)
	\nn \\
	&&
	        \times
		w'(E_{i-1},E_i,E_{i+1})
		e^{E_{i+1}/k_BT_c}
		e^{E_i/k_BT_c}
		e^{E_{i-1}/k_BT_c}
	\nn\\		
	&=&
		\int_{-\infty}^{0}d\left(\frac{E_{i}}{k_BT_c}\right)
		2\nu_{1}e^{E_i/k_BT}
		e^{-2\nu_{1}e^{E_i/k_BT}t}
		e^{E_i/k_BT_c}
	\nn\\	
	&&+
		\int_{-\infty}^{0}d\left(\frac{E_{i}}{k_BT_c}\right)
		\nu_{1}e^{E_i/k_BT}
		e^{-\nu_{1}e^{E_i/k_BT}t}
		e^{E_i/k_BT_c}
	\nn\\&&\times
		\int_{-\infty}^{0_-}d\left(\frac{E_{i-1}}{k_BT_c}\right)
		e^{E_{i-1}/k_BT_c}
	\nn
	\\	
	&&+
		\int_{-\infty}^{0}d\left(\frac{E_{i}}{k_BT_c}\right)
		\nu_{1}e^{E_i/k_BT}
		e^{-\nu_{1}e^{E_i/k_BT}t}
		e^{E_i/k_BT_c}
	\nn\\&&
		\times\int_{-\infty}^{0_-}d\left(\frac{E_{i+1}}{k_BT_c}\right)
		e^{E_{i+1}/k_BT_c}.
		\label{11080501}
\end{eqnarray}
Here, $0_-\equiv 0-\lim_{\delta\to+0}\delta$.
Note that, due to the introduction of a specific energy $E_c(=0)$, the integral in Eq. (\ref{wtev}) should be 
performed independently for the three sites $i, i\pm 1$, in the region $E_i, E_{i\pm 1} \in (-\infty, E_c)$, respectively.
Now, taking the lmit $t\gg\tau_1$, and using Eq. (\ref{11072902}), we have
\begin{eqnarray}
	&&
		\langle w'(E_{i-1},E_i,E_{i+1})\rangle
	\nn\\
	&&\simeq
		\nu_{1}\frac{T}{T_c}(2\nu_{1}t)^{-1-\frac{T}{T_c}}\Gamma\left(\frac{T}{T_c}+1\right)
	\nn\\&&
		+
		2\nu_{1}\frac{T}{T_c}(\nu_{1}t)^{-1-\frac{T}{T_c}}\Gamma\left(\frac{T}{T_c}+1\right)
		\lim_{\delta\to0}e^{\delta}
	\nn\\&&
		\propto t^{-1-\frac{T}{T_c}}.
		\label{11080502}
\end{eqnarray}
From Eq. (\ref{11080502}), we have $\alpha=T/T_c$. 
This is exactly the result of the MTM \cite{TMMA,MTM}. 
The generalization of the above result to the system with arbitrary 
numbers of neighbouring sites is tedious but straightforward
, and the same result can be obtained. 
Thus, the MTM can be expressed as a special case of the disorder model.
From the viewpoint of the disorder model, the spatial factor $N_R(t)$ in the
anomalous exponent is exactly cancelled by the constraints imposed above, leaving $\alpha$ to
be superficially insensitive to the spatial structure.

\section{Summary}
In this study, we have calculated the anomalous exponent in the disorder model.
First, we have shown that the anomalous exponent is time-dependent. 
Furthermore, since the characteristic time scales $\tau_n$ where the $n$
th-nearest neighbours contribute to the effective waiting time are well
separated, i.e., $\tau_{n+1} \gg \tau_n$, 
the $(n+1)$ th-nearest neighbours can be ignored for time scales around
or below $\tau_n$. 
This indicates that the number of the effective neighbouring sites
${N}_{R}(t)$ is finite in realistic (experimental) time scales.
In fact, we have shown that it is enough to consider 
at most the third-nearest neighbours for typical systems. 
On the other hand, since ${N}_{R}(t)$ increases monotonically with time,
$\alpha(t)$ eventually becomes $1$ at some time in the future, i.e., 
subdiffusive transport eventually becomes normal.
This is consistent with the common knowledge of nonequilibrium statistical mechanics.

Secondly, we have shown that the {\it spatial structure} of the sites might be a novel and
 an equally significant element in the disorder model, in addition to the two common elements, the temperature $T$
and the typical width of the DOS, $T_c$.
Specifically, the {\it network structure} of the sites is shown to be the novel element, which can be seen in the expression
for the anomalous exponent, $\alpha(t) = N_R(t) T/T_c$, where $N_R(t)$ is
 the number of the ``effective neighbouring sites'' at time $t$. 
This leads to the feature that other elements, such as the distance between the sites or the spatial dimension of the system, 
are insensitive to the time-dependence of the electric transport, as long as $N_R(t)$ is identical.

We have verified the above theoretical results by means of MC simulation of the hopping conductance.
First, we chose a one-dimensional system with carriers allowed to hop up
to the second-nearest neighbouring sites, to verify the transition form subdiffusive 
to normal transport. 
The result actually showed the transition, and 
the anomalous exponent was identical to the theoretical values.
Next, we chose a one-dimensional system with carriers allowed to hop up
 to the third-nearest neighbouring sites, to verify that 
$(n+1)$ th-nearest neighbours can be ignored for time scales around or below $\tau_n$. 
The result was consistent with the theoretical prediction.
Moreover, plateaus were found around each $\tau_n$. 
This indicates that the anomalous exponent can be regard as constant
for certain time intervals.
This is compatible with the common treatment where the anomalous
exponent is assumed to be constant.
Finally, a one-dimensional and a two-dimensional system, with identical
network structures, were compared, to verify that the {\it network structure} determines the anomalous exponent.
The results were identical for the two cases, in agreement with the
asymptotic results for the anomalous exponent, Eq. (\ref{anom-exp-N3}).

We have also revealed the relation between the disorder model and the MTM \cite{TMMA,MTM}, 
which was not stated and remained obscure before.
We have shown that the MTM can be expressed as a constrained version of
the disorder model, where only hopping to the energy levels of the conduction band
is allowed.
This constraint exactly cancels the factor $N_R(t)$ in the expression of
 the anomalous exponent.
Whether the system is better described by the ``general
(unconstrained)`` disorder model or 
the ``constrained'' disorder model (i.e., MTM)
 is a matter of the nature of the system, and 
 hence expected to be case dependent. 

{\it Acknowledgements}.
We are grateful to Mr. Shinjo, Dr. Okuda, and the members of the Analysis Technology Development Department 1 for 
fruitful discussions and 
their support.
\bibliography{EgamiSuzukiWatanabe}
\end{document}